\documentclass[doublecol]{epl2} 
\usepackage{epsf}
\usepackage{amsfonts}
\usepackage{amssymb}
\usepackage{amsthm}
\usepackage{graphicx}

\newcommand{\cf}[1]{\langle #1 \rangle}                      
\newcommand{\no}[1]{: \! #1 \! :}                            
\newcommand{\vJ}{\mbox{\boldmath $J$}}                       
\newcommand{\vS}{\mbox{\boldmath $S$}}                       
\newcommand{\phdagger}{\mathop{\phantom{\dagger}}}           
\newcommand{\psiop}[1]{\psi^{\phdagger}_{#1}}                
\newcommand{\psidop}[1]{\psi^{\dagger}_{#1}}                 
\newcommand{\Psiop}[1]{\Psi^{\phdagger}_{#1}}                
\newcommand{\aop}[1]{a^{\phdagger}_{#1}}                     
\newcommand{\adop}[1]{a^{\dagger}_{#1}}                      

\newcommand{\vsigma}{\mbox{\boldmath $\sigma$}}

\newcommand{\bml}{\begin{mathletters}}
\newcommand{\eml}{\end{mathletters} \hspace{-5pt}}

\title{Enhanced two-channel Kondo physics in a quantum box device}

\author{Paata Kakashvili\inst{1,2} \and Henrik Johannesson\inst{3}}
\shortauthor{P. Kakashvili and H. Johannesson}

\institute{                    
  \inst{1} Department of Applied Physics, Chalmers University of Technology,
SE-412 96 G\"oteborg, Sweden\\
  \inst{2} Department of Physics and Astronomy, Rice University, Houston, TX
  77005, USA\\
  \inst{3} Department of Physics, G\"oteborg University,
  SE-412 96 G\"oteborg, Sweden
}

\abstract{We propose a design for a one-dimensional quantum box device where
the charge fluctuations are described by an anisotropic two-channel Kondo model.
The device consists of a quantum box in the Coulomb blockade regime,
weakly coupled to a quantum wire by a single-mode point contact. The
electron correlations in the wire produce strong back scattering at
the contact, significantly increasing the Kondo temperature as
compared to the case of non-interacting electrons. By employing
boundary conformal field theory techniques we show that the
differential capacitance of the box exhibits manifest two-channel
Kondo scaling with temperature and gate voltage, uncontaminated by the
one-dimensional electron correlations. We discuss the prospect to
experimentally access the Kondo regime with this type of device.}

\begin{document}

\maketitle

The study of the Kondo effect has been at the forefront of condensed
matter research ever since its inception forty years ago~\cite{FortyYears}.
While the simplest case studied by Kondo $-$ a spin-1/2 impurity coupled
to a single band of conduction electrons $-$ is by now well
understood, a number of variations of the original problem continue
to challenge the experimentalist as well as the theorist.

A particularly intriguing question is how to realize the
(overscreened) {\em two-channel Kondo effect} in an
experiment. Ideally, two-channel Kondo physics emerges when there are
two competing channels in which the conduction electrons can screen a
spin-1/2 impurity. As a result, the impurity spin becomes overscreened
below some characteristic temperature $T_K$, and various thermodynamic
and transport properties show non-Fermi liquid (NFL) behavior~\cite{AD,TW,AL}.
Being the simplest example of NFL behavior driven by
a localized degree of freedom, the model that encapsulates the effect
$-$ the two-channel Kondo model~\cite{NB} $-$ has attracted enormous
interest. Many experimental realizations have been suggested over the years, 
including a proposal for a quantum dot device where a small
spinful dot (the ``impurity'') is coupled to two screening channels
defined by a pair of conducting leads and a larger dot~\cite{OregGoldhaberGordon}.
A major challenge is how to control that 
the couplings to the two channels are identical $-$ as required for the
two-channel Kondo effect to appear$-$ and also, how to make sure that there
is no channel mixing from cotunneling of electrons (which would immediately destroy
the effect). It was recently
reported that the required level of control may in fact be achieved in the
laboratory,
and some experimental data in support of two-channel Kondo physics was
presented~\cite{Potok}. 

A different approach is to search for realizations of two-channel
Kondo physics in systems where the channel symmetry and independence
are robustly protected by some conservation law. A case in point is a quantum
box (a large semiconducting quantum dot or a metallic grain) weakly
connected by a point contact to a conducting lead. As shown by
Matveev~\cite{Matveev}, near a degeneracy point of the average charge of the
box the charge fluctuations can be modeled by an anisotropic
two-channel Kondo Hamiltonian. The two available charge states in the
box (corresponding to $n-1$ and $n$ electrons) take the role of the
two spin states of the impurity, while the physical spin of the
conduction electrons provide for the two independent channels.
In the absence of a magnetic field, this guarantees that the channel
symmetry is robust. Unfortunately, the Kondo temperature $T_K$ of this device is very
small, $T_K \sim E_C\mbox{e}^{-1/2t \nu}$, $E_C$ being the single-electron charging
energy of the box, and $t$ the tunneling rate through the
contact. Trying to increase $T_K$ by increasing $E_C$ requires that
the box is made smaller, which in turn threatens to kill the effect
since $T_K$ must still be larger than the level spacing in the box.
Whereas great progress has been achieved in high-precision charge
measurements \cite{Lehnert}, and fingerprints of two-channel 
Kondo charge-fluctuations may
have been identified in the capacitance of a semiconducting quantum
box connected to a lead~\cite{Berman}, the difficulties to satisfy
the conflicting constraints above make it unlikely that a fully developed
effect can be observed in such a device~\cite{Zarand}.

In this Letter we consider a novel design for a quantum box device
with an enhanced Kondo temperature, allowing for a possible
experimental entry to two-channel Kondo physics. Our scheme, inspired
by that of Matveev~\cite{Matveev}, adds to several recent proposals 
for realizing the two-channel Kondo effect in a nanoscale
structure~\cite{GramespacherMatveev, KimKimKallin, LeHurSeelig,
LebanonSchillerAnders, ShahMillis, LeHurSimon, LeHurSimonBorda, FlorensRosch,
AndersLebanonSchiller, Pustilnik, BolechShah}. With our effort we
also wish to address an issue that has been notably absent from 
the discussion of this problem: the possible influence from {\em
dynamic} electron interactions on a two-channel charge Kondo
effect. 

\begin{figure}[!hpb]
\begin{center}
\includegraphics[width=3.4in]{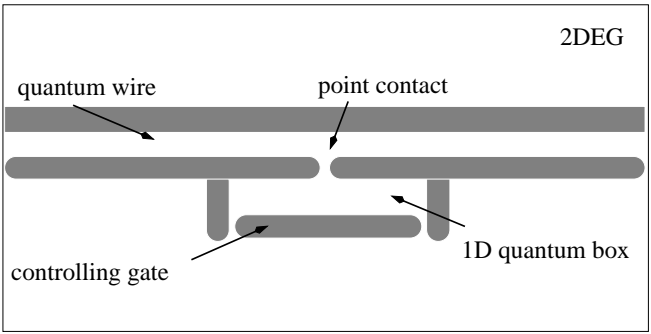}
\caption{Schematic picture of the proposed setup. A 1D quantum box side-coupled
  to a quantum wire via a  point contact.}
\label{setup}
\end{center}
\end{figure}

Our design, which is most easily implemented in a gated
semiconductor heterostructure or cleaved edge overgrowth
structure~\cite{Auslaender}, is sketched in fig.~\ref{setup}: A one-dimensional (1D)
quantum box is side-coupled to a quantum wire via a low-transmission
single-mode point contact, putting the box in the Coulomb blockade regime.
The box is biased by a gate voltage $V$, initially tuned to a value
where the charging energy for $n-1$ and $n$ electrons is the same
{\em(degeneracy point)}: $V=-ne/2C_{\Sigma}$ with $n$ an odd integer,
$e$ the electron charge, and $C_{\Sigma}$ the capacitance of the
box. The box should be made sufficiently large so that its level
spacing $\Delta$ is much smaller than any other energy scale in the
problem, allowing for the discrete levels to be represented by a quasi-continuum.

Introducing second-quantized operators $\aop{k \mu \alpha}$ for
electrons in the wire ($\alpha=0$) and the box ($\alpha=1$)
with momentum $k$ and spin $\mu=\uparrow, \downarrow$, we model the
set-up by the Hamiltonian
\begin{equation} \label{Hamiltonian}
H=H_{0}+H_{I}+H_{C}+H_{T}\,,
\end{equation}
where 
\begin{eqnarray}
H_{0}&=&\sum_{k, \mu, \alpha} \epsilon^{\phdagger}_k \adop{k \mu \alpha}
\aop{k \mu \alpha}\,,  \nonumber \\
H_I&=&\frac{1}{2}\sum_{\alpha, \alpha'\atop \mu, \mu'} \sum_{k, k', q} V_{\alpha
\alpha'}(q)\adop{k+q \,\mu \alpha}
\adop{k'-q\,\mu' \alpha'}\aop{k' \mu' \alpha'} \aop{k \mu \alpha}\,, \nonumber \\
H_{C}&=&\frac{Q^{2}}{2C^{}_{\Sigma}}+ VQ\,, \label{Hc} \nonumber \\
H_{T}&=&t\sum_{k,k',\mu}(\adop{k\mu 0}
\aop{k' \mu 1} + h.c.)\,, \nonumber \label{Htun}
\end{eqnarray}
The term $H_{C}$ encodes the charging energy of the box with $Q$ the surplus charge
with respect to the zero-bias Fermi level, while $H_{T}$ governs the tunneling through
the point contact, $t$ being a constant tunneling matrix element.  
By construction we have split the electron interaction into two pieces:
The {\em mean-field} capacitive part $H_{C}$ that is effective only in
the finite box (assuming that the charging energy of the wire can be
neglected), and a part $H_{I}$ which builds dynamic electron
correlations into the model~\cite{footnote1}. $H_{I}$ is most easily
specified in the relevant low-energy limit where all scattering
processes are confined to the neighborhood of the Fermi points $\pm
k_F$. Assuming that the electron density is incommensurate with the
lattice of the underlying substrate, the allowed low-energy processes
can be classified into dispersive, forward and backward scattering 
(with the latter taking place only inside the wire or box since there is
no exchange of wire and box electrons away from the point contact).
Passing to a continuum description and decomposing the electron fields
$\Psiop{\mu \alpha}(x) \sim \int dk \mbox{e}^{ikx}
\aop{k\mu\alpha}$ in left- and right-moving parts, $\psiop{-\mu \alpha}(x)$ and
$\psiop{+ \mu \alpha}(x)$, respectively, these processes can be conveniently
expressed using the currents 
\begin{eqnarray} \label{Currents}
J_{\pm}&=& \mbox{sh}\vartheta \no{\psidop{\pm \mu \alpha}\psiop{\pm \mu \alpha}}+
\mbox{ch}\vartheta \no{\psidop{\mp \mu \alpha}\psiop{\mp \mu \alpha}}\,,
\nonumber \\
\vJ_{\pm}^{[\mu]}&=&\frac{1}{2}
\no{\psidop{\pm \mu \alpha}{\vsigma}_{\mu \mu'} \psiop{\pm \mu' \alpha}}\,, \\
\vJ_{\pm}^{[\alpha]}&=&\frac{1}{2}
\no{\psidop{\pm \mu \alpha}{\vsigma}_{\alpha \alpha'} \psiop{\pm \mu \alpha'}}.
\nonumber
\end{eqnarray}
The normal ordering is taken w.r.t. the filled Dirac sea (obtained after linearizing
the spectrum
around $\pm k_F$), $\vsigma$ is the vector
of Pauli matrices, and the indices $\alpha, \alpha', \sigma, \sigma'$ are summed over.
The parameter $\vartheta$ is given by $2\vartheta = \mbox{arctanh}(3g/(v_F + 3g))$,
with
$v_F$ the Fermi velocity and with $g$ the strength of the screened Coulomb
interaction, 
here approximated by its dominating component at zero-momentum transfer.
For simplicity we take this
interaction to be the same {\em between} wire and box ($\sim V_{01}(q)$) as {\em in}
the wire and the box 
($\sim V_{00}(q)=V_{11}(q)$):
$V_{01}(0)=V_{00}(0)\equiv g$. In a real device one expects that
$V_{01}(0)<V_{00}(0)$. 
However, as will be seen below, the value of
the ratio $V_{01}(0)/V_{00}(0)$ at most influences subleading corrections to the
charge fluctuations in the box,
and for the purpose of extracting the leading behavior we may put it to unity,
yielding a more transparent
formalism.
Given the currents in (\ref{Currents}) we can now cast the low-energy limit of
$H_0+H_I$ in the form~\cite{Senechal}
\begin{eqnarray}    \label{Sugawara}
H_0 + H_I &=& \frac{1}{2\pi} \sum_{\ell =\pm \atop \eta = \alpha, \mu} \int  
\upd x\, ( \frac{v_c}{8} \no{J_{\ell}(x) J_{\ell}(x)}  \nonumber \\ 
&+&\frac{v_{[\eta]}}{4}\no{\vJ_{\ell}^{[\eta]}(x) \cdot
\vJ_{\ell}^{[\eta]}(x)} ) \,,
\end{eqnarray}
with $v_{c}=v_{F}(1+6g/v_{F})^{1/2}, v_{[\mu]} = v_{[\alpha]} = v_F-g$. 
We have here removed two marginally irrelevant interactions, including an unphysical
exchange process
between wire and dot electrons away from the point contact \cite{footnote2}.
The U(1) charge, SU(2)$_2$ spin and SU(2)$_2$ {\em pesudospin} currents $J_{\ell}, 
{\vJ}_{\ell}^{[\mu]}$ and ${\vJ}_{\ell}^{[\alpha]}$, respectively, are decoupled in
(\ref{Sugawara}), and
one recognizes $H_0+H_I$ as the Hamiltonian for a spinful Luttinger 
liquid (written in ``Sugawara form''~\cite{Senechal}) with an extra pseudospin
channel. 
Having thus coded the dynamic part of the theory we turn to its
effect on the charge fluctuations in the box.  

In the vicinity of the degeneracy point and with $k_BT \ll e^2/2C_{\Sigma}$, only
the $Q=0$ and $Q=e$ states
are accessible and higher charge states can be removed from the theory. Following
Matveev~\cite{Matveev}, the 
resulting two-level system $H_C + H_T$
in (\ref{Hamiltonian}) can then be mapped onto an anisotropic two-channel Kondo
interaction $H_K$:
\begin{equation}  \label{KondoInteraction}
H_{K}=\frac{J_{\perp}}{2}
\psidop{\ell  \mu \alpha}(0)
{\vsigma}^j_{\alpha \alpha'}\psiop{\ell' \mu \alpha'}(0) S^j - hS^z\,.
\end{equation}
Here $\vS$ is a pseudospin-1/2 operator that implements the constraint on the
allowed charge states
in the box. The coupling $J_{\perp}$ and the field $h$ are given by  
$J_{\perp}=2t$ and $h=eu$ respectively,  with $u$ a small voltage bias away from the
degeneracy point. 
Note that {\em all} indices in (\ref{KondoInteraction}),
($\ell, \ell' = \pm ;\, \alpha, \alpha' = 0,1; \mu, \mu' = \uparrow,
\downarrow , j = x,y)$, 
are summed over. To complete the mapping to a Kondo pseudospin formulation one makes
use of the fact that
$\cf{Q(u)}=e[1/2-\cf{S^z}(h,J_{\perp})]$, implying that the differential capacitance
$c(u,T)$ of the box gets modeled
by an impurity pseudospin susceptibility $\chi_{imp}(h,T) \equiv -(1/e^2)\partial
\cf{Q}/\partial u = c(u,T)$.
Having translated the original problem into Kondo language, the task has thus become
that of calculating the 
susceptibility of a pesudospin-1/2
impurity (eq.~(\ref{KondoInteraction})) coupled to a two-channel Lutttinger liquid
$H_0+H_I$ (eq.~(\ref{Sugawara})), 
with the two channels provided by 
the physical spin of the electrons. 

It is important to realize that the backscattering of electrons off the impurity
in (\ref{KondoInteraction}) is due to the side-coupling of the box to
the wire, see fig.~\ref{setup}. As it turns out, it is precisely this novel
feature that yields the desired properties of our set-up. As shown by
Le Hur, the amplitude for two-channel electron-impurity backscattering
renormalizes to a strong coupling fixed point as the temperature is
lowered~\cite{LeHur}. This is in exact analogy with the single-channel
Kondo problem in a Luttinger liquid, where this effect was first 
noted~\cite{LeeToner}. The flow to strong coupling produces a crossover from
an exponentially suppressed Kondo temperature $T_K \sim
E_c\mbox{e}^{-1/2t\nu}$ for $g \ll J_{\perp}$ to a power law $T_K \sim
E_C(2t\nu)^{2/(1-K_c)}$ for $g \gg J_{\perp}$, with $K_c$ the
Luttinger liquid charge parameter. Exploiting the crossover formula from
Ref.~\cite{LeeToner}, with input parameters chosen for a GaAs
based device~\cite{Goni} and taking $t \nu \lesssim 0.2$, one finds that
while $T_K$ in the noninteracting limit $(K_c = 1)$ is comparable to or below the
level spacing in the box, $T_K$ becomes almost an order of magnitude
larger as $K_c$ approaches 0.6 (which is easily reached in a
low-density quantum wire~\cite{EXP}). Although the temperature window
that opens is probably too narrow for a full-fledged two-channel Kondo
scaling to develop, it should at least allow for a controlled experimental
study of its transient behavior. Using a {\em metallic}  
quantum wire/box with its much larger effective electron mass would stretch the
window by another order of magnitude (provided that the electron density
is suppressed by proper gating of the device). We here point to the recent observation
that electron-shell effects can stabilize arbitrarily long metallic quantum wires,
making the fabrication of a metallic device a viable and realistic prospect~\cite{BurkiStafford}.

Suppose that the limitations set by current semiconductor based technology can indeed be
overcome, allowing for the capacitance to be measured in the critical
region $T\ll T_K$ using a metallic device. Would a logarithmic scaling with temperature and
gate voltage emerge $-$as predicted for noninteracting electrons~\cite{Matveev}$-$
or will the strong dynamic electron correlations in the
wire and the box cause a new type of behavior? In the case of spin polarized
electrons it has been shown that 1D correlations strongly influence
charge fluctuations~\cite{KJ}, and one may anticipate a similar
outcome also in the present case. To find out, we have employed the
tools of boundary conformal field theory (BCFT)~\cite{DiFrancesco},
building on an earlier approach to the isotropic two-channel Kondo effect
in a Luttinger liquid~\cite{GJ}. The required analysis closely follows
that presented in ref.~\cite{GJ}, and we here 
only sketch the key ideas together with the
results. 

Let us first recall that the fixed point of the {\em isotropic} two-channel 
Kondo effect in a Luttinger liquid corresponds to a particular
selection rule for quantum numbers of the BCFT embedding
$\otimes_{i=1,2}$[U(1)$\otimes$SU(2)$_{2}$$\otimes$SU(2)$_2$]$^i$
implied by the Hamiltonian in (\ref{Sugawara})
\cite{GJ}. Here the U(1) factor represents charge,
while the SU(2)$_2$ factors represent spin and pseudospin, with $i=1,2$
corresponding to left and right moving fields. The Kondo interaction couples
left and right movers, and therefore the symmetries above are broken
down to their diagonal subgroups. For the charge sector this implies
that any U(1) operator with dimension
\begin{equation}
\Delta_{c}\!=\!\frac{1}{4}n^{2}e^{\pm 2
  \theta}\!+\!N; \hspace{1cm} n,N \in \mathbb{N}
\end{equation}
is allowed \cite{GJ}.
Factorizing the diagonal subgroups in the spin and pseudospin sectors  
amounts to a coset construction at the level of conformal towers. The
two SU(2)$_{2}$ towers in the spin and pseudospin sectors are
decomposed into SU(2)$_{4}$ and a coset which is generated by the N=1
superconformal algebra (SCA) of central charge c=1:
SU(2)$_{2} \otimes$SU(2)$_{2}=$SU(2)$_{4} \otimes$SCA.
Primary states of
the spin (pseudospin) SU(2)$_{4}$ sectors have conformal dimensions
$j(j+1)/6$ with $j \in \{0,1/2,1,3/2,2\}$. The SCA in turn is divided
into two sectors: the Ramond (R) and Neveu-Schwartz (NS) algebras with
primary dimensions $\{1/24,1/16,3/8,9/16 \}$ and $\{ 0,1/16,1/6,1 \}$
respectively. In addition, the grade of a generic state is integer in
the R sector, whereas it is half-integer in the NS sector.
The effect of the Kondo coupling is encoded by the 
{\em leading irrelevant boundary operators} \cite{AL} 
(which are products of operators from charge, spin, pseudospin and SCA sectors)
with possible scaling dimensions
\begin{equation}
\Delta=\Delta_{c}+\sum_{j=s,\,p}(\Delta^{j}_{SU(2)_{4}}+\Delta^{j}_{SCA}),
\end{equation}
where
\begin{eqnarray}
\Delta^{s/p}_{SU(2)_{4}}&\!\!=\!\!& \frac{j(j+1)}{6}\!+\!N, \hspace{0.5cm} j=
0,\frac{1}{2},1,\frac{3}{2},2; \\ 
\Delta^{s/p}_{SCA}&\!\!=\!\!& \{0,\frac{1}{16},\frac{1}{6},1\}\!+\!\frac{N}{2},~
 \{\frac{1}{24},\frac{1}{16},\frac{3}{8},\frac{9}{16}\}\!+\!N,
\end{eqnarray}
with $N \in \mathbb{N}$. Valid boundary operators (i) must respect 
all symmetries of the theory
and (ii) must not violate the known critical scaling of observables in
the non-interacting limit $g \rightarrow 0$. The criterion (i)
restricts the choice of operators in each conformal sector, while
the criterion (ii) defines the rules for combination of operators in
different sectors to obtain the full boundary operators.

Given the above construction and criteria it is in principle
straightforward to pinpoint the effect from the exchange anisotropy
(broken pseudospin rotational symmetry) and the magnetic field (broken
time reversal symmetry) in (\ref{KondoInteraction}). In contrast to
the SU(2)$_{4}$ invariant case above where only operators that
transform as singlets are allowed, more operators now appear in the
pseudospin sector. By inspection we find that the only relevant
operator produced is the $j=1$ primary field $\phi^{z}$, with
conformal dimension $\Delta_{\phi^{z}}=1/3$. This operator is present only
if pseudospin rotational {\em and} time reversal symmetries are
simultaneously broken. In the limit of vanishing magnetic field the
anisotropy is irrelevant, implying that the magnetic field is a relevant
perturbation, as for the two-channel Kondo model for non-interacting
electrons \cite{Pang}. 

There are two more operators appearing because
of the broken pseudospin symmetry: The first descendant $J^{z}$ of the $j=0$
identity operator, and the $j=2$ primary field $\phi^{zz}$, both being
exactly marginal of dimension $\Delta=1$. Both operators may be combined
with others from charge, spin, and SCA sectors to form new composite
operators provided that these respect the constraints (i) and (ii) above.
For the {\em isotropic} problem the leading behavior of the
impurity susceptibility $\chi(T,h)$ is driven by the same operator (of
dimension $\Delta = 3/2$) as for noninteracting electrons, giving rise
to a logarithmic divergence as $T\rightarrow 0$ or $h \rightarrow 0$
\cite{GJ}. In our construction the boundary operator driving the
behavior is a combination of the first descendant of the pseudospin
j=1 SU(2)$_{4}$ conformal tower with the $\Delta=1/6$ NS field. To explore 
whether a faster divergence may result from any of the new composite
operators generated from the broken pseudospin symmetry, we have to
identify those of scaling dimension $\Delta < 3/2$ (since they
may produce more leading contributions) and then test them
against criteria (i) and (ii) above.

We can here identify two distinct classes of possible boundary operators. The first
class contains operators with dimensions $\Delta<3/2$, which do
not depend on the Luttinger liquid interaction parameter $K_{c}$ (thus
containing the identity operator from the charge sector) and hence
they should be present in the non-interacting limit as well. Since these
contributions do not vanish in the non-interacting limit, the 
constraint (ii) implies that the operators from this
class are not present in the theory. The second
class contains operators with dimensions $\Delta(K_{c})<3/2$,
which depend on the Luttinger liquid interaction parameter $K_{c}$ (thus
containing non-trivial operators from the charge sector). A straightforward
calculation, using the method described in Ref. \cite{GJ}, reveals 
that the contributions to the observables from these operators do
not vanish in the non-interacting limit. Thus, again using constraint (ii),
we conclude that these operators should also not be included in 
the theory.

It follows that the leading behavior of the
differential capacitance $c(T,u)$ of our proposed set-up exhibits the same
logarithmic scaling as in the two-channel Kondo effect for
noninteracting electrons,
\begin{equation}  \label{temperaturescalingII}
c(T,u\!=\!0) = A \ln \left( \frac{T_K}{T} \right)+ \cdots, \ \ \ T \ll T_K
\end{equation}
and
\begin{equation} \label{voltagescalingII}
c(T\!=\!0,u)= B \ln \left( \frac{T_H}{eu} \right)+ \cdots, \ \ \ eu \ll T_H
\end{equation} 
but {\em with significantly larger Kondo temperatures $T_K$ and
$T_H \approx T_K$.}
Here $A$ and $B$ are
constants, and ''$\cdots$'' indicate subleading terms.
These subleading terms $-$ produced by 
more irrelevant operators than those
identified above $-$ may turn out to be interaction dependent.
To determine their scaling dimensions, however, requires a cumbersome
analysis, beyond the scope of the present Letter.
It is here interesting to compare with the single-channel case
where already the {\em leading} scaling behavior depends on the strength
of the electron-electron interaction \cite{KJ}.
 
In conclusion, we have shown that charge fluctuations close to a
degeneracy point of a 1D Coulomb blockaded quantum box side-coupled to
a quantum wire exhibit logarithmic two-channel Kondo divergences with
temperature $T$ (voltage $u$) for $T \ll T_K$ ($eu \ll T_H$). This leading
behavior is {\em not} modified by the strong 1D electron correlations in the
wire and the box (with possible interaction dependent contributions at most
showing up in subleading terms). The Kondo temperature $T_K$ (or $T_H$ in the case
of voltage scaling)
can be significantly larger compared
to a device with non-interacting electrons.  While design constraints
for a semiconductor implementation probably only allow for crossover
effects to be observed, the fabrication of 
a metallic device should yield access to the full two-channel charge
Kondo effect. 

\acknowledgments
We thank S. Eggert and M. Granath for helpful discussions. This work was
supported by the Swedish Research Council under grant number 621-2002-4947.


\begin{thebibliography}{0}
\bibitem{footnote1}
For simplicity, in $H_{0}$ and $H_{I}$ we
take the length of the box to be equal to that of the wire,
putting all finite-size effects into $H_{C}$.
\bibitem{footnote2}
The excluded interactions can be written in terms of $SU(4)_{1}$ Kac-Moody currents
as $H_{excl}=-2g J^{{\cal A}}_{L}J^{{\cal A}}_{R}$ (${\cal A}=1,...,15$),
where $H_{excl}$ is marginally irrelevant for $g>0$.
\end{thebibliography}
\end{document}